# PubMed and Beyond: Biomedical Literature Search in the Age of Artificial Intelligence

Qiao Jin, Robert Leaman, Zhiyong Lu
National Center for Biotechnology Information
National Library of Medicine
National Institutes of Health
{qiao.jin, robert.leaman, zhiyong.lu}@nih.gov

## Abstract

Biomedical research yields a wealth of information, much of which is only accessible through the literature. Consequently, literature search is an essential tool for building on prior knowledge in clinical and biomedical research. Although recent improvements in artificial intelligence have expanded functionality beyond keyword-based search, these advances may be unfamiliar to clinicians and researchers. In response, we present a survey of literature search tools tailored to both general and specific information needs in biomedicine, with the objective of helping readers efficiently fulfill their information needs. We first examine the widely used PubMed search engine, discussing recent improvements and continued challenges. We then describe literature search tools catering to five specific information needs: 1. Identifying high-quality clinical research for evidence-based medicine. 2. Retrieving gene-related information for precision medicine and genomics. 3. Searching by meaning, including natural language questions. 4. Locating related articles with literature recommendation. 5. Mining literature to discover associations between concepts such as diseases and genetic variants. Additionally, we cover practical considerations and best practices for choosing and using these tools. Finally, we provide a perspective on the future of literature search engines, considering recent breakthroughs in large language models such as ChatGPT. In summary, our survey provides a comprehensive view of biomedical literature search functionalities with 36 publicly available tools.

# Introduction

In biomedicine, literature serves as the primary means of disseminating new findings and knowledge. Much of the information accumulated by biomedical research remains accessible only through the literature[1]. Consequently, literature search, the process of retrieving scientific articles to satisfy specific information needs, is integral to all aspects of biomedical research and patient care. For example, to practice evidence-based medicine, clinicians must locate relevant literature that depicts similar scenarios[2]. Similarly, for knowledge discovery, biomedical researchers rely on insights from prior publications as new knowledge often builds on prior knowledge[3].

The exponential growth of biomedical literature makes identifying the information relevant to a given information need challenging. PubMed, the most widely used biomedical literature search engine, currently contains nearly 36 million articles, with the addition of more than 1 million annually. A typical PubMed query retrieves hundreds to thousands of articles, yet fewer than 20% of the articles past the top 20 results are ever reviewed[4,5]. This motivated a shift in PubMed's approach from recency-based ranking to a relevance-based ranking[6], to better prioritize the most relevant and significant articles.

However, PubMed primarily serves as a general-purpose biomedical literature search engine. Despite significant improvements over the past decades[5], PubMed mainly processes short keyword-based queries, returning a list of raw articles without further analysis. Consequently, it might not optimally serve specialized information needs, which require alternative query types or have specific requirements for ranking articles or displaying results. A notable example is the unprecedented upsurge of publications addressing the COVID-19 pandemic[7,8]. While the pandemic made quickly disseminating new findings critical, obtaining comprehensive results from traditional search engines requires complex querying syntax that is unfamiliar to most users and manual topic indexing requires several months post-publication. Addressing the COVID-19 pandemic, therefore, required a specialized literature search engine capable of automatically collecting and classifying relevant articles[9,10].

While various web-based literature search tools have been proposed over the past two decades to complement PubMed for specific literature search needs, they remain underutilized and unfamiliar to clinicians and researchers. This survey aims to acquaint readers with available tools, discuss best practices, identify functionality gaps for different search scenarios, and ultimately facilitate biomedical literature retrieval. Literature search tools included in this study meet the following criteria: they must be web-based, freely available, regularly maintained, and designed for searching the biomedical literature.

Consequently, general-domain literature search engines such as Web of Science, Scopus, Google Scholar, and Semantic Scholar, are not included.

Table 1 enumerates the web-based literature search tools introduced in this survey, categorized by the unique information needs they fulfill. Specifically, literature search tools are classified into five areas: (1) Evidence-based medicine (EBM), for identifying high-quality clinical evidence; (2) Precision medicine (PM) and genomics, for retrieving information related to genes or variants; (3) Semantic search, for finding textual units semantically related to the input query; (4) Literature recommendation, for suggesting related articles; and (5) Literature mining, for extracting biomedical concepts and their relations for literature-based discovery. Figure 1 presents a high-level overview of the search scenarios. Search tools catering to different information needs differ in the types of queries they accept, their methods for processing articles and matching them to the input query, and how they present search results to users.

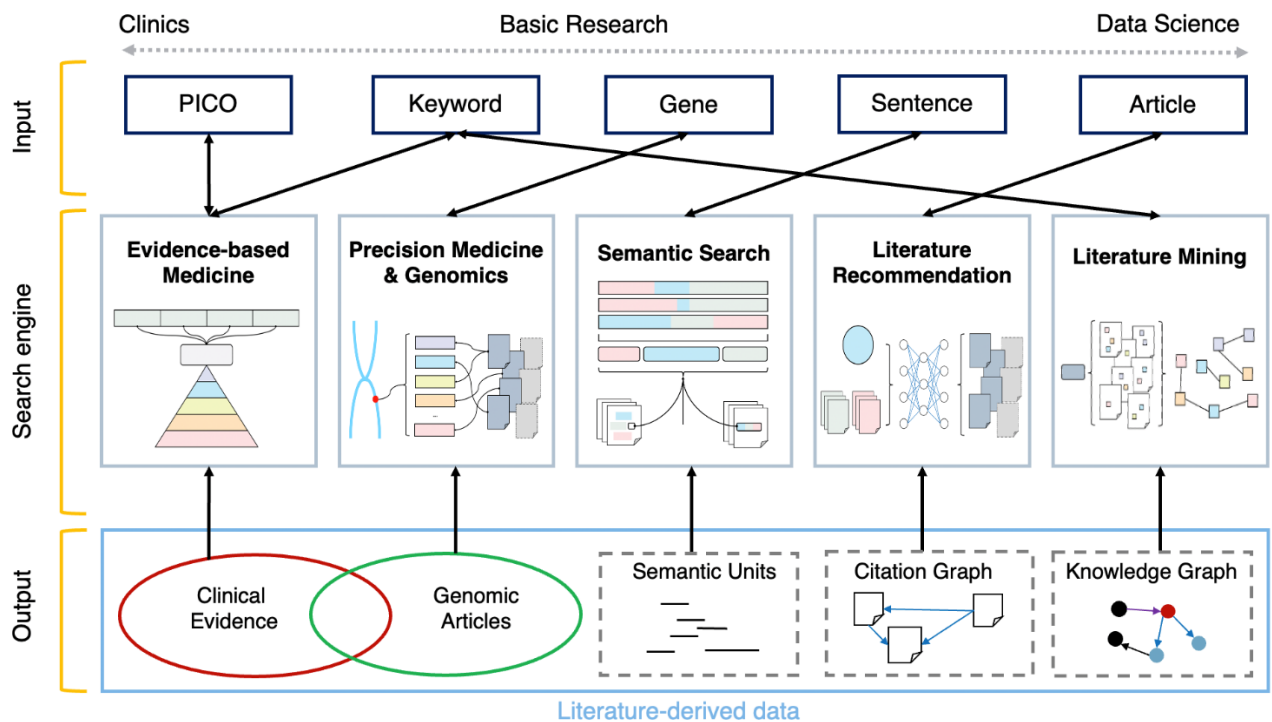

**Figure 1**. Overview of five specialized search scenarios in biomedicine: evidence-based medicine, precision medicine & genomics, semantic search, literature recommendation, and literature mining. Each search scenario is characterized by its unique input interface, search or ranking algorithm, and output display.

This article differs from previous surveys on biomedical literature search tools[11-14] in three important aspects: (1) We organize the literature search tools into a novel, scenario-oriented, classification system, providing users with a straightforward and instructive guide for selecting the most suitable tool for their information needs; (2) Our study includes many newly-introduced systems not covered by previous surveys; (3) Beyond surveying current systems, we also cover practical considerations and best practices of choosing and using these tools for addressing biomedical information needs. Finally, we share our perspective on the development of next-generation biomedical literature search engines. Specifically, we discuss how large language models (LLM), such as ChatGPT, could be integrated with traditional literature retrieval to iteratively refine information needs into queries, retrieve relevant articles, then summarize the information retrieved or provide answers directly. Our goal is to provide a comprehensive overview of specialized literature search tools for clinicians and researchers to consider for different use cases, which enables more effective exploration of biomedical information and higher-quality care for their patients.

**Table 1**. Web-based biomedical literature search tools. Literature search tools included in this study are web-based, freely available, regularly maintained, and designed for searching the biomedical literature.

| **Resource** | **Website** | **Brief description** |
| --- | --- | --- |
| **General** | | |
| PubMed | https://pubmed.ncbi.nlm.nih.gov/ | General-purpose biomedical literature search engine. |
| PubMed Central | https://www.ncbi.nlm.nih.gov/pmc/ | Supporting full-text search. |
| Europe PMC | https://europepmc.org/ | Searching both abstracts and full-texts. |
| **Evidence-based Medicine (EBM)** | | |
| PubMed PICO Search | https://pubmedhh.nlm.nih.gov/pico/index.php | Searching clinical studies with PICO elements. |
| PubMed Clinical Queries | https://pubmed.ncbi.nlm.nih.gov/clinical/ | Searching clinical studies with various type and scope filters. |
| Cochrane Library | https://www.cochranelibrary.com/ | Searching high-quality systematic reviews. |
| Trip Database | https://www.tripdatabase.com/ | General EBM search engine. |
| **Precision Medicine (PM) and Genomics** | | |
| LitVar | https://www.ncbi.nlm.nih.gov/research/litvar | |

| Variant2literature | https://www.taigenomics.com/console/v2l | Searching relevant information for all synonyms to the given variant. |
|---|---|---|
| DigSee | http://210.107.182.61/geneSearch/ | Finding evidence sentences for the given (gene, disease, biological processes) triplet. |
| OncoSearch | http://oncosearch.biopathway.org/ | Searching sentences that mention gene expression changes in cancers |
| **Semantic Search** | | |
| LitSense | https://www.ncbi.nlm.nih.gov/research/litsense/ | Searching relevant sentences to the given query. |
| COVID-19 Challenges and Directions | https://challenges.apps.allenai.org/ | Searching COVID-19 challenges and future directions for the given topic. |
| askMEDLINE | https://pubmedhh.nlm.nih.gov/ask/index.php | Answering the query question with documents or text snippets in literature. |
| COVID-19 Research Explorer | https://covid19-research-explorer.appspot.com/biomedexplorer/ | Answering the original question and follow-up questions with text snippets in literature |
| BioMed Explorer | https://sites.research.google/biomedexplorer/ | |
| **Literature Recommendation** | | |
| LitCovid | https://www.ncbi.nlm.nih.gov/research/coronavirus/ | Literature hubs for COVID-19. |
| WHO COVID-19 Research Database | https://www.who.int/emergencies/diseases/novel-coronavirus-2019/global-research-on-novel-coronavirus-2019-ncov | |
| iSearch COVID-19 Portfolio | https://icite.od.nih.gov/covid19/search/ | |
| Corona Central | https://coronacentral.ai/ | |
| LitSuggest | https://www.ncbi.nlm.nih.gov/research/litsuggest/ | Scoring article candidates based on user-provided |

| BioReader | https://services.healthtech.dtu.dk/service.php?BioReader-1.2 | positive and negative articles. |
|---|---|---|
| Connected Papers | https://www.connectedpapers.com/ | Recommending relevant articles to one or more seed articles using the citation graph. |
| Litmaps | https://www.litmaps.com/ | |
| **Literature Mining** | | |
| PubTator | https://www.ncbi.nlm.nih.gov/research/pubtator/ | Highlighting biomedical concepts in the retrieved documents. |
| Anne O'Tate | http://arrowsmith.psych.uic.edu/cgi-bin/arrowsmith_uic/AnneOTate.cgi | Ranking the extracted concepts from the search results. |
| FACTA+ | http://www.nactem.ac.uk/facta/index.html | Finding directly and indirectly associated concepts to the given concept. |
| BEST | http://best.korea.ac.kr/ | Displaying graphs of biomedical concepts and their relations extracted from the retrieved documents. |
| Semantic MEDLINE | https://ii.nlm.nih.gov/SemMed/semmed.html | |
| SciSight | https://scisight.apps.allenai.org/ | |
| PubMedKB | https://www.pubmedkb.cc/ | |
| LION LBD | https://lbd.lionproject.net/ | |
| **Experimental literature search systems augmented by large language models (LLMs)** | | |
| Scite | https://hippocratic-medical-questions.herokuapp.com/ | Finding relevant articles to users' question and then using LLMs to answer the question with the retrieved articles |
| Elicit | https://elicit.org/ | |
| Consensus | https://consensus.app/ | |
| Statpearls semantic search | https://hippocratic-medical-questions.herokuapp.com/ | |

## PubMed & PMC: the first stop

PubMed, a widely used search engine for biomedical literature, is developed and maintained by the US National Library of Medicine (NLM). In 2021, it averaged approximately 2.5 million queries daily. When a query is entered, PubMed expands it to include additional Medical Subject Headings (MeSH) terms[1] via automatic term mapping. The search engine then seeks exact matches for this expanded query in the indexed fields of each article, including the title, abstract, author list, keywords, and MeSH terms. Traditionally, all matching articles were returned in reverse chronological order. A new machine-learning-based retrieval model – Best Match – was introduced in 2017 to better assist users by returning the most relevant articles among the top results[6]. Best Match considers multiple ranking signals such as an article's type, age, and usage, and is trained with past user search logs. As of 2020, Best Match has become the default sorting option in the new PubMed[15]. Beyond relevance search for biomedical topics, PubMed also supports various other search functionalities. These include matching single citations[2] through bibliographic information such as title and journal names, as well as Boolean operators that are usually used when conducting systematic reviews.

However, since PubMed does not index full-text articles, those that match the query in the full-text but not in the abstract or the title will not be retrieved. Such queries are accommodated by PubMed Central (PMC), which provides access to more than 7 million freely available full-text articles. Unfortunately, PMC does not support searching the other 27 million PubMed articles that lack full-text availability. Europe PMC[16], a PMC partner, contains both 42.7 million abstracts and 9.0 million full-text articles as of July 2023.

While PubMed and PMC might be an ideal starting point for keyword queries, their utility beyond keyword-based searches is limited. For example, Shariff et al. demonstrate that the results for unfiltered PubMed queries are much less efficient and comprehensive than filtered queries when retrieving clinical evidence[17]. Allot et al. report that searches for genomic variants such as "rs121913527" often return zero results in PubMed[18] despite synonymous variants being mentioned in many articles. Additionally, Fiorini et al. find that queries exceeding five words tend to retrieve less satisfactory results in PubMed[5]. These findings suggest a need for specialized search engines to meet specific information needs.

**Best Practice:** PubMed should be the first choice for three types of literature search practices: (1) exploring biomedical topics via keyword query, with PMC enabling keyword search within the full text, when available; (2) searching for a single citation, i.e., a specific article; (3) reproducible literature screening with Boolean queries.

---

[1] https://www.nlm.nih.gov/mesh/meshhome.html
[2] https://pubmed.ncbi.nlm.nih.gov/citmatch/

# Evidence-based medicine

Evidence-based medicine (EBM)[19] requests clinical practitioners follow high-quality evidence, primarily derived from peer-reviewed articles of clinical studies. Efficient retrieval of this evidence is crucial for implementing EBM. Accordingly, clinical questions should be structured effectively, incorporating at least the "PICO" elements[20] (**P**opulation, **I**ntervention, **C**omparison, and **O**utcome). For example, in "Does remdesivir reduce in-hospital mortality for COVID-19 patients compared to placebo?", the PICO elements are COVID-19 patients (Population), remdesivir (Intervention), placebo (Comparison), and in-hospital mortality (Outcome), respectively. EBM search engines should be equipped to process both PICO and natural language clinical questions.

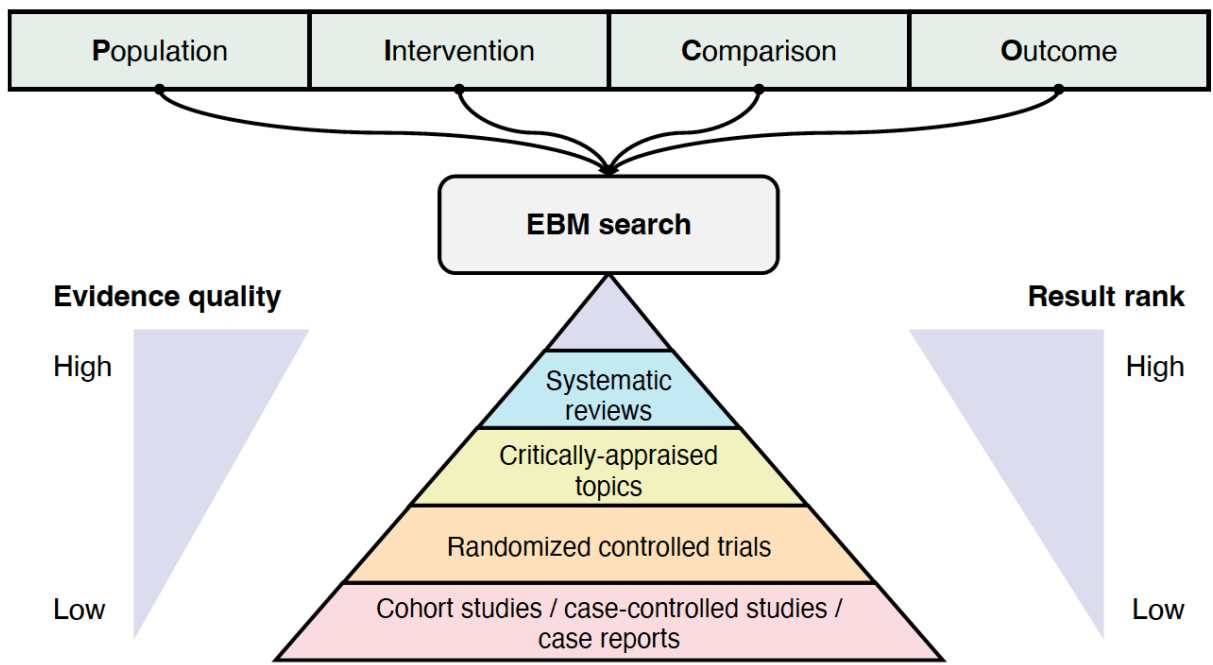

**Figure 2.** The architecture of a search engine for evidence-based medicine (EBM). EBM search engines should incorporate PICO elements (Population, Intervention, Comparison, and Outcome) within the input query and rank the articles returned based on the quality of the evidence.

Clinical evidence spans a broad spectrum of literature, with significant variability in quality. For example, systematic reviews are generally considered as higher-quality evidence than randomized controlled trials (RCTs), which, in turn represent higher quality than individual case reports. Consequently, an ideal EBM search engine should consider the quality of evidence for filtering or ranking the articles. Figure 2 depicts the architecture of an ideal EBM search engine, which allows PICO-style input and ranks results based on evidence quality.

**Systems accepting PICO queries:** Several EBM search engines, such as Trip Database, the Cochrane PICO search, and Embase, accommodate PICO-based queries. PubMed for Handhelds[21], a lightweight platform designed for handheld devices in clinical settings, also supports PICO-based search. The search interfaces for these systems typically contain text boxes corresponding the four primary PICO elements. In general, these systems provide more precise results since the search intent is explicitly stated in the query. For example, entering "diabetes" as the "Population" term, prompts EBM search engines to only return clinical studies on diabetes patients. In contrast, keyword-based search engines would return any article that mentions "diabetes," regardless of its relevance to patient studies.

**Systems with filtered retrieval results:** PubMed Clinical Queries search employs predefined filters[22,23] for clinically-relevant studies of various types, such as therapy and diagnosis. Users can also select broad (general) or narrow (specific) scopes for the filters. Clinical practitioners should use the narrow scope for a quick overview of the important studies at the point of care, while researchers synthesizing evidence should employ the broad scope for exhaustive searches. Several EBM search engines prioritize retrieval of secondary evidence, such as systematic reviews and critically-appraised topics, which typically have higher quality than primary evidence. A notable example is the Cochrane Database, which hosts over 11 thousand high-quality systematic reviews and protocols. Critically-appraised topics summarize the evidence on a specific topic, such as prevention of type 2 diabetes mellitus, using short, templated, titles to simplify retrieval. As a result, they provide convenient clinical decision support in systems like UpToDate, a commercial evidence-based clinical resource[3].

**Assisting evidence synthesis:** Compared to evidence retrieval, fewer systems facilitate evidence synthesis, the systematic collection, analysis and combination of results from multiple research studies to reach a comprehensive conclusion about a specific question or topic[24]. Evidence synthesis culminates in the creation of high-quality publications such as systematic reviews. However, the user conducting an evidence synthesis would need to manually screen all related literature to address a clinical question without bias, an extremely time-consuming process due to the vast number of articles likely to be relevant across multiple databases[25]. Despite efforts to use machine learning to automate this screening process[26], these features are not yet integrated into web-based EBM search engines.

---

[3] https://www.wolterskluwer.com/en/solutions/uptodate

**Best practice**: Literature search is a vital step in evidence-based medicine. To optimize this process, users should: (1) formulate clinical questions in the format of PICO elements; (2) utilize a system that ranks relevant studies by their evidence quality.

## Precision medicine and genomics

Precision medicine (PM) is an emerging approach that tailors disease treatment and prevention based on individual variations in genes, environment, and lifestyle[27]. The rapid development of high-throughput sequencing techniques have precipitated a sharp decline in the cost of obtaining individual genomic data, surpassing the reduction predicted by Moore's Law[28]. Human genomes, with their high heterogeneity, contain a large number of genomic variants. It is estimated, for example, that there are 4 to 5 million single-nucleotide polymorphisms (SNPs) in a person's genome[29]. Understanding the biological function and clinical significance of these genomic variants is essential for the advancement of precision medicine.

These data are typically stored in manually curated databases[30] such as UniProt[31], dbSNP[32], ClinVar[33], and Gene Ontology[34]. These databases manually summarize and maintain primary findings from the literature about each data entry (e.g., variant, gene, or protein). However, the growth of the biomedical literature, with an average of 3,000 new articles per day[1], outpaces the speed of manual curation, leaving a knowledge gap. To supplement these databases, search engines capable of extracting gene or variant-related information directly from raw literature are needed. This section primarily discusses such systems.

A significant challenge for PM and genomics search engines is the presence of multiple representations or synonyms for the same genomic variant. For instance, the variant "V600E" could also be referred to as "Val600Glu," "1799T>A," or "rs113488022." This synonymy causes retrieval challenges for keyword-based search engines. In response, many specialized literature retrieval tools have been proposed; their core functionality is shown in Figure 3, where the search engine should be able to retrieve all articles that mention the exact variant query as well as its synonyms.

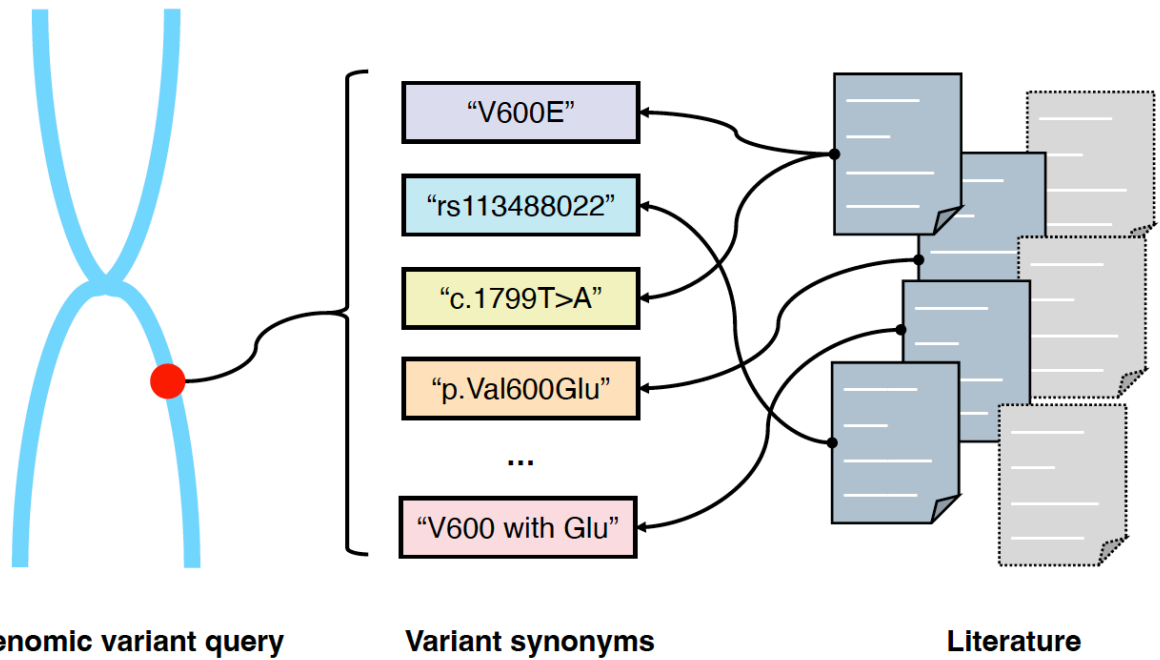

**Figure 3.** Illustration of the functionality of a search engine for precision medicine and genomics. Search engines for precision medicine and genomics should handle queries containing genomic variants and identify all synonymous references to these variants in the literature.

**Recognizing synonymous mentions:** Some tools, such as LitVar[18,35], focus on normalizing variant synonyms in the literature. LitVar uses text mining tool tmVar[36,37] to recognize variant names and convert them to standardized form. LitVar indexes both abstracts from PubMed and full-texts from PubMed Central and is updated regularly to ensure retrieval of all current literature containing synonyms of the query. Another tool, variant2literature[38], provides a structured query interface that allows users to specify a chromosome location . Unique to  variant2literature is the ability to extracts variants from figures and tables in addition to the article text.

**Linking genes and other information:** Several systems go beyond recognizing synonymous gene mentions and explore genomic-related information. DigSee[39] accepts a triplet of gene, disease, and biological processes as input and finds sentences in PubMed abstracts that link the gene to the disease through the given biological processes. Evaluation studies have shown that DigSee covers more gene-disease relations than manually curated databases like Online Mendelian Inheritance in Man[40], and that the findings are also reliable[41]. OncoSearch[42] specializes in retrieving literature evidence for gene expression changes and cancer progression status. Specifically, it annotates sentences from the literature to indicate whether the input gene is up-regulated or down-regulated, whether the input cancer progresses or regresses with the expression change, and the expected role of the gene in the cancer. Other tools such as DISEASES[43] and DisGeNet[44] offer a search interface to gene-disease relations extracted from literature but do not return the raw evidence.

**Best practice:** To find relevant information about a gene or a variant, we recommend first querying curated databases such as UniProt and ClinVar. For more recent findings or when these databases lack sufficient data to contextualize a specific variant, the use of search engines specialized for precision medicine and genomics is recommended. For example, LitVar can assist in finding information within the literature about the role of certain genomic variants in an emerging disease, which might not have been curated into structured databases yet.

## Semantic search

Unlike the keyword-based search that seek exact matches for the input query, semantic search locates texts that are semantically related to the query. For example, "renal" and "kidney" are semantically very similar. Figure 4 outlines semantic search, where text units such as concepts and sentences that match the query semantically are returned, such as mentioning the same diseases and discussing possible treatments. These texts do not necessarily contain the exact terms from the query, making their retrieval by traditional literature search engines unlikely. While there are various forms of semantic relevance, semantic search engines typically focus on one type. We introduce search engines for two common types of semantic relevance: similar sentences and question-answer pairs.

**Similar sentence search:** Article-level searches often overlook finer-grained information in sentences. Sentence-level searches are important for precise knowledge retrieval. For example, one can search for a particular finding and compare it with relevant findings from other articles. LitSense[45], a web-based system for sentence retrieval from PubMed and PMC, utilizes an embedding-based retrieval system in addition to traditional keyword matching. Results in LitSense can be filtered by sections, such as Conclusions. While LitSense searches for all types of similar sentences, several literature search engines have also been proposed for more specific types of sentences[46-48]. Lahav et al. present a search engine for sentences that describe challenges and future directions in COVID-19[46]. SciRide Finder[47] finds cited statements describing the in-line references. BioNOT[48] indexes and searches sentences that contain negated events. While separate semantic search engines for different sentence types can provide valuable functionality, a unified similar search engine integrated with filters provides greater flexibility and a consistent user interface.

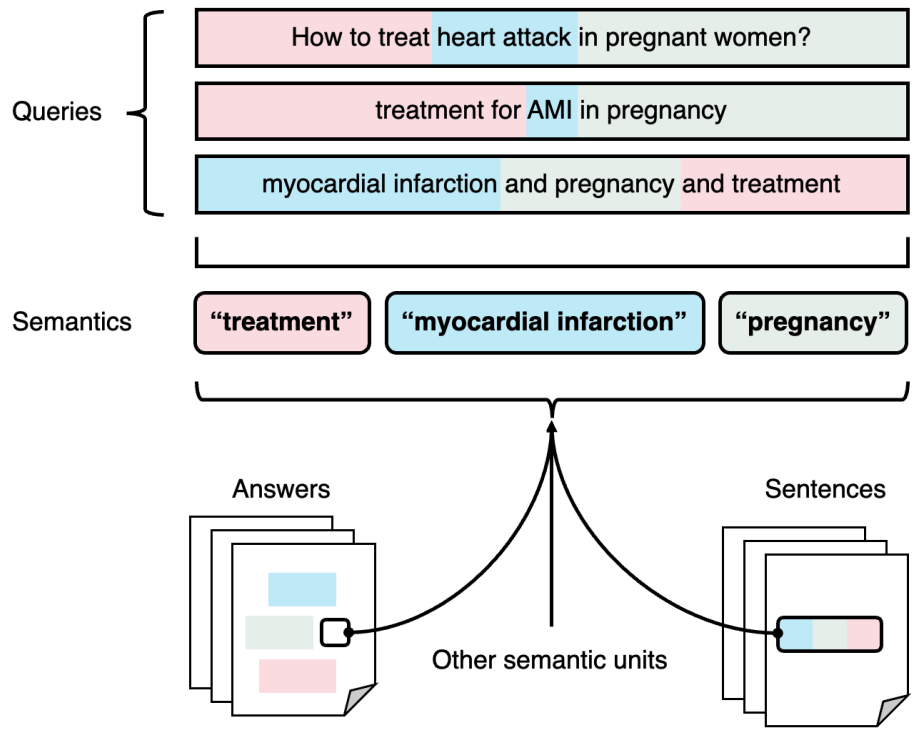

**Figure 4.** Depiction of semantic search. Unlike traditional keyword-based search engines, semantic search engines process words and phrases according to their meaning rather than the literal text. For instance, "heart attack", "AMI", and "myocardial infarction" share similar meanings.

**Question answering:** Biomedical inquiries are often naturally expressed as questions, such as the PICO-based clinical questions in EBM. However, traditional keyword-based search engines may not efficiently handle natural language questions because questions and answers often lack high lexical overlap. Biomedical question answering (QA) is an active research area[49], but user-friendly web tools remain sparse. The askMEDLINE[50] system evolved from PubMed PICO search and enables direct input to the clinical questions, e.g., "Is irrigation with tap water an effective way to clean simple laceration before suturing?". askMEDLINE displays results as a list of relevant articles. AskHERMES[51], a clinical QA system, performs semantic analysis on complex questions and extracts summaries from the relevant articles to directly answer the question, which is more convenient than a list of relevant articles which must be searched. COVID-19 Research Explorer and BioMed Explorer are experimental semantic search engines for biomedical literature developed by Google AI. The former focuses on COVID-19 articles in CORD-19[52], and the latter encompasses all PubMed articles. Both explorers are presumably based on Google's systems[53,54] designed for BioASQ[55], a challenge for biomedical information retrieval and QA, and have a modern search interface. Users ask natural language questions, and the answers are highlighted in the text snippets in the results. Users can also pose follow-up questions to further investigate the research topic.

**Best practice:** Users should consider using semantic search engines if their information needs are better expressed by natural language instead of keywords. Available tools include LitSense for finding relevant sentences and BioMed Explorer for answering biomedical questions with evidence from the literature.

## Literature recommendation

Biomedical research often requires comprehensive exploration of related literature. Traditional keyword-based search engines are typically inefficient for this purpose due to the difficulty of formulating queries to exhaustively capture all relevant work. Literature recommendation engines instead allow users to explore articles relevant to a specific research topic or similar to a list of articles known to be relevant. This section mainly introduces two types of literature recommendation tools: topic-based and article-based, as depicted in Figure 5. Additional forms, such as passage-based literature recommenders or citation recommenders[56] are still experimental and have not been implemented as web applications.

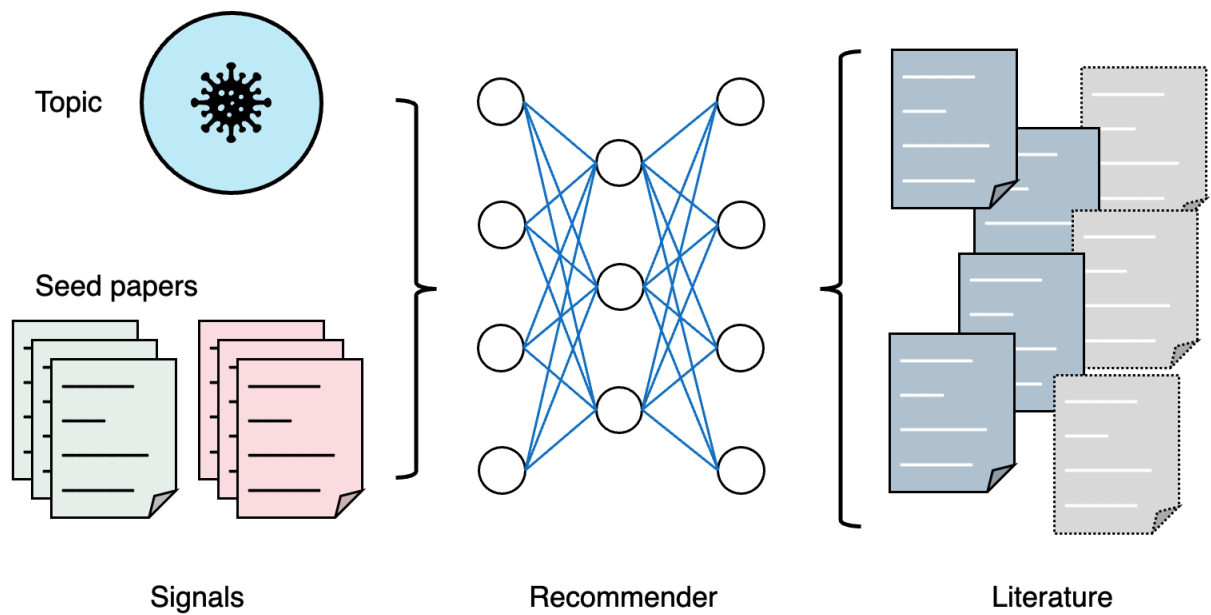

**Figure 5.** Illustration of topic-based and article-based literature recommendation systems. Topic-based systems provide articles relevant to a specific topic, while article-based systems return articles similar to a group of initial (seed) articles and dissimilar to a group of irrelevant articles.

**Topic-based literature recommendation systems** are typically curated databases or literature hubs tailored to selected research topics, such as the COVID-19 pandemic. For example, due to the initial lack standardized terminology for SARS-CoV-2 and COVID-19, publications used a variety of terms, complicating identifying relevant articles through keyword-based or Boolean searches. LitCovid[10,57], a curated literature hub containing COVID-19-related articles from PubMed, is organized with eight broad topics, including mechanism, transmission, diagnosis, and treatment. Chen et al. demonstrated that LitCovid identifies about 30% more PubMed articles than a complex, purpose-built Boolean query[10]. Other literature hubs dedicated to COVID-19 include the WHO COVID-19 Research Database[4], the iSearch COVID-19 portfolio[5], CoronaCentral[58] and etc.

**Article-based literature recommendation systems**, on the other hand, generate a list of articles related to one or more initial (seed) articles. Modern literature search engines often provide a list of articles related to individual articles, such as the "similar articles" section in PubMed[59]. A few systems have been proposed, however, which support identifying articles related to a list of articles instead of individual ones. LitSuggest[60], a literature recommendation system based on machine learning, rates candidate articles on their similarity to a user-supplied list of positive articles and dissimilarity to a list of negative articles. The list of negative articles is optional, with random articles used if not supplied by the user. Users can also provide human-in-the-loop feedback by annotating a subset of the scored candidate articles and re-training the recommendation model. BioReader[61] offers similar functionality, but it requires a list of negative articles. Several commercial literature search tools like Connected Papers[6] and Litmaps[7] provide visual representations of articles related to seed articles on a citation graph, thus aiding in the navigation of the academic literature and guiding focused research. A significant shortcoming of current systems is their inability to explain their recommendations[62], which is particularly important in high-stakes fields like biomedicine.

**Best practice:** Recommendation systems primarily assist in literature exploration. Users can find articles related to a topic of interest, such as COVID-19, using a curated literature database, or locate articles similar to a specific list of articles through article-based literature recommenders like LitSuggest.

---

[4] https://www.who.int/emergencies/diseases/novel-coronavirus-2019/global-research-on-novel-coronavirus-2019-ncov
[5] https://icite.od.nih.gov/covid19/search/
[6] https://www.connectedpapers.com/
[7] https://www.litmaps.com/

# Literature mining

Literature mining aims to uncover novel insights from scientific publications through natural language processing (NLP) techniques[63]. NLP tasks include named entity recognition (NER), the task of recognizing named entities (biomedical concepts) such as genes and diseases[64], and relation extraction (RE), which classifies relations between the entities identified[65]. For example, an NER tool could identify a genetic variant and a disease name in a sentence, and an RE tool might classify their relation as mutation-causing-disease. Extracted concepts and their relations can be organized into a graph, referred to as a knowledge graph, which structurally summarizes the knowledge encoded in the publications related to the given query. By displaying a knowledge graph, literature search engines provide users with an overview of the knowledge discovered, thereby facilitating new knowledge discovery by predicting potential missing links. This process is visualized in Figure 6.

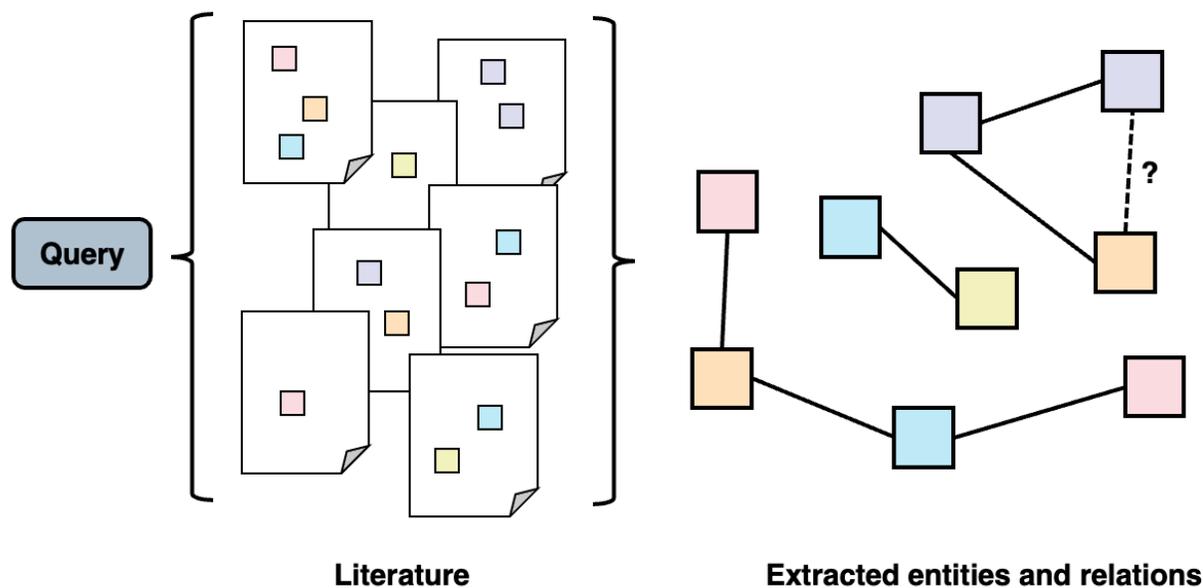

**Figure 6.** The architecture of a system for mining entity associations from the biomedical literature. The system retrieves articles relevant to a given query, extracts biomedical entities and their relationships (e.g., variant-causing-disease), and presents the search as a knowledge graph that visualizes the extracted entities and their relationships.

**Entity-augmented literature search:** Several literature search engines enhance the retrieved results with biomedical concepts. PubTator[66,67] highlights six types of concepts recognized by state-of-the-art NER tools, including genes, diseases, chemicals, mutations, species, and cell lines. PubTator has also made its annotations publicly available via bulk download and an application programming inference, allowing other search engines to augment the search results with PubTator concepts. Notably, PubTator has been integrated into platforms such as LitVar, LitSense, and LitCovid. Anne O'Tate[68,69] provides options to rank concepts, such as important words, important phrases, topics, authors, MeSH pairs, etc., that are extracted from the retrieved articles.

**Relation-augmented literature search:** Some systems further process the extracted concepts and show the search results using associated concepts. FACTA+[70] finds concepts associated with the given concept and the supporting sentences and can uncover indirectly associated concepts through certain types of "pivot concepts" as the bridge. BEST[71] is a biomedical concept search tool that returns a list of biomedical concepts, including genes, diseases, targets, and transcription factors. Evidence sentences for concept relevance are also displayed in BEST. Semantic MEDLINE[72] extracts SemRep[73] predications, comprising two Unified Medical Language System[74] (UMLS) concepts and one UMLS semantic relation, from the retrieved articles and provides a graph visualization of the predications. SciSight[75], an exploratory search system for COVID-19, can present a graph of biomedical concepts associated with the given concept. PubMedKB[76] extracts and visualizes semantic relations between variants, genes, diseases, and chemicals, offering a user interface with interactive semantic graphs for the input query. The LION literature-based discovery system[77] also presents the search results as a graph that contains biomedical concepts and their relations extracted from the literature. While many systems for constructing biomedical knowledge graphs automatically have been proposed, there has been less research on how these systems can assist users in literature-based knowledge discovery. The utility of knowledge graphs in this context remains to be confirmed in future studies.

**Best practice**: Literature mining tools can be employed to study the associations between biomedical concepts in the literature. Users should consider the concept and relation types of interest and choose the literature mining tools that incorporate such information. For example, PubTator provides annotations for six general concept types, but concepts beyond these types and concept relations are better supported in other literature search tools, such as SciSight for COVID-19 concepts and relations.

# Looking Ahead: The Role of ChatGPT and Other Large Language Models in Literature Search

ChatGPT[78] and other large language models (LLMs) such as PaLM[79] have recently demonstrated considerable performance improvements on both general and biomedical domain-specific NLP tasks. There is a rising belief that these models could significantly change how users interact with information online, potentially including the biomedical literature[80]. In this section, we explore the potential application of LLMs to the biomedical literature search scenarios presented in this review.

**Evidence-based medicine**: LLMs can accelerate evidence synthesis in two key ways. First, they can suggest Boolean queries to aid literature screening for systematic reviews[81]. Following the retrieval of results, LLMs could potentially be used to summarize and synthesize the resulting articles[82,83]. However, preliminary evaluations have exposed various issues which must be addressed before widespread use. Apart from evidence synthesis, LLMs can also enhance the extraction of PICO elements from the medical literature[84], thereby improving PICO-based EBM search engines.

**Precision medicine and Genomics**: Most genomics information resides in curated databases, which are not easily accessible due to their keyword-centric search functions and less modern user interfaces. LLMs can alleviate these access difficulties by autonomously utilizing tools such as the application programming interface (API) of specialized databases[85].

**Semantic search**: LLMs have achieved state-of-the-art performance on several biomedical QA datasets[86,87], which require clinical knowledge and biomedical reasoning capabilities[88,89]. This suggests that LLMs can provide direct responses to users' natural language questions using relevant documents returned from a traditional search engine. This feature, called retrieval augmentation, is already supported by experimental literature search engines such as scite[8], Elicit[9], and statpearls semantic search[10]. These tools accept a natural language research question as input and retrieve relevant articles through semantic search. They further prompt LLMs such as GPT-3[90] to answer the user's question based on the retrieved relevant articles. The systems return both the relevant articles and the LLM-generated answers. This is commonly known as retrieval augmentation and has also been integrated into general web search engines such as the new Bing. However, these LLM-generated answers are susceptible to precision and recall errors[91-93] and should be carefully verified before use.

**Literature recommendation:** The potential role of LLMs in literature recommendation remains largely unexplored. One possibility involves using LLMs to explain literature recommendations, i.e., describing why a recommended article is similar to the input article. This capability could be used to create a dataset for training smaller generative

---

[8] https://scite.ai/
[9] https://elicit.org/
[10] https://hippocratic-medical-questions.herokuapp.com/

models, enabling more flexible and cost-effective and literature recommendation explanations.

**Literature mining**: Unlike other literature search scenarios that benefit from the generative capabilities of LLMs (e.g., summarization for semantic search), literature mining predominantly depends on NLP tasks such as NER and RE. LLMs generally do not outperform smaller task-specific models such as BERT[94], fine-tuned specifically for these tasks[95]. However, LLMs may offer superior interpretations of the constructed knowledge graphs, revealing previously unknown associations between biomedical concepts.

# Discussion

In this article, we introduced five specific use cases of biomedical literature search and available tools for each scenario. Two of these scenarios are application-oriented (EBM, PM), while three are technique-oriented (semantic search, literature recommendation and mining). Our classification, while practical, is not mutually exclusive, and the advantages of different systems can be combined to better meet diverse biomedical information needs. For instance, an EBM search engine might also process queries where the specified Population is associated with certain genomic variants, necessitating recognition of variant synonyms for comprehensive literature retrieval. Another instance is biocuration, the practice of converting literature data into database entries. This three-step process[96], involves document selection, indexing documents with biomedical concepts, and extracting their specific relations or interactions. A system to support biocuration should be equipped with both literature recommendation and mining functionality to assist biocurators by suggesting relevant publications and highlighting the relevant biomedical concepts.

Users might be expected to prefer a single search portal that can fulfill multiple information needs. As current literature search systems are specialized and scattered, we believe a unified portal integrating all specific functionalities would facilitate access to biomedical literature. At a high-level, biomedical literature search systems comprise three key components, as depicted in Figure 1: the input interface, the ranking algorithm, and the output display. The search systems we have introduced differ primarily in these components, which we discuss below.

**Input interface:** Analogous to web search, literature search queries generally comprise several words[4,5]. Consequently, most literature search engines accept short text inputs, typically representing biomedical concepts or concepts, such as an author's name or a disease. In this review, we broadly denote such systems as keyword-based. However, more complex or specialized information needs require interfaces capable of processing semi-structured information or even non-text modalities. Semi-structured search interfaces accept separate texts for multiple pre-defined fields, akin to the advanced search interface in modern literature search engines and PICO-based EBM search. Some information needs defy expression in text, such as finding articles that are similar to one set but dissimilar to another set of articles, requiring interfaces designed specifically for the task. Although modern search interfaces consisting of one text box are simple and easy to use, the resulting queries can be ambiguous or overly general. As a result, task-oriented search interfaces should be designed for different biomedical literature search purpose, while a unified portal can be employed to triage the user's information needs into these task-oriented interfaces.

**Ranking algorithms:** In literature search engines, the ranking algorithms assess article relevance for a given query, thereby determining which articles are returned to the user. PubMed employs the Best Match[6] ranking model, a machine learning approach trained via user click logs. Many other ranking algorithms, such as the BM25 algorithm[97], rank articles based on the importance of the terms which overlap between the article and the query. These algorithms calculate general text-based relevance without domain-specific requirements, while certain biomedical subdomains have specific article ranking requirements. For example, in EBM, articles with higher quality clinical evidence should be ranked higher. In semantic search, articles with text units that are semantically related to the input query should be returned, irrespective of term overlap. For article-based literature recommendation, the system should rank articles similar to the positive seed articles and dissimilar to the negative seed articles. In addition to performing purpose-specific ranking, future literature search engines should incorporate transparent and interpretable ranking algorithms.

**Output display:** search results are most commonly displayed as a list of article metadata, including titles, authors, publication types and dates, abstract snippets, and so forth, mimicking the general web search engines familiar to users. Though list-based display has been almost unchanged in general search engines for two decades, additional modules have been introduced to serve specific information needs. For example, many web search engines directly display the answer to a question query at the top of the results page, mirroring the goal of QA-based semantic search in biomedical literature. Certain literature mining systems construct and visualize a knowledge graph from the articles retrieved, aiding exploration and knowledge discovery. Given the remarkable text generation capabilities of LLMs such as ChatGPT, we anticipate future literature search engines will include high-level overviews of returned articles generated by LLMs.

## Conclusion

Our aim has been to assist biomedical researchers and clinicians in finding the most suitable literature search tool to fulfill their specialized information needs. Specialized search engines may serve specific information needs in the biomedical literature more effectively than general-purpose systems. We characterized search scenarios for five specific information needs: evidence-based medicine, precision medicine and genomics, semantic search, literature recommendation, and literature mining. We included 36 systems for biomedical literature search and classified each according to the unique information needs they fulfill. All tools discussed are web-based, freely available, regularly maintained, and designed for searching the biomedical literature. Finally, we discussed the future of biomedical literature search, especially considering the potential impacts of large language models (LLMs) such as ChatGPT.

## Acknowledgment

This research was supported by the NIH Intramural Research Program, National Library of Medicine.